\documentclass[a4paper,12pt]{article}

\usepackage{amssymb}
\usepackage{graphicx}
\usepackage[english]{babel}

\usepackage{amsmath}
\usepackage{cleveref}
\usepackage{multirow}
\usepackage{tabularx}
\usepackage{xcolor}
\usepackage{url}
\usepackage{mathtools}

\urldef{\mailsc}\path|{cristina.fernandez, carlos.vela, merce.villanueva}@uab.cat|

\newtheorem{theorem}{Theorem}[section]
\newtheorem{corollary}{Corollary}[section]
\newtheorem{proposition}{Proposition}[section]
\newtheorem{lemma}{Lemma}[section]
\newtheorem{example}{Example}[section]

\newenvironment{proof}[1][Proof.]{\vspace{0.5em}\textbf{#1} }{\
\hfill$\square$}

\newcommand{\Z}{\mathbb{Z}}
\newcommand{\zero}{{\mathbf{0}}}
\newcommand{\one}{{\mathbf{1}}}
\newcommand{\two}{{\mathbf{2}}}
\newcommand{\C}{{\cal C}}

\newcommand{\wt}{{\rm wt}}
\newcommand{\rank}{\text{rank}}
\newcommand{\kernel}{\text{ker}}

\newcommand{\cS}{{\cal S}}

\begin{document}

\title{On the Kernel of $\Z_{2^s}$-Linear Simplex and MacDonald Codes
\thanks{This work has been partially supported by the Spanish MINECO under Grant TIN2016-77918-P
(AEI/FEDER, UE), and by the Catalan AGAUR under Grant
2017SGR-00463. The authors are with the Department of Information and Communications
Engineering, Universitat Aut\`{o}noma de Barcelona, 08193 Cerdanyola del Vall\`{e}s, Spain.}}

\author{Cristina Fern\'andez-C\'ordoba, Carlos Vela, Merc\`e Villanueva}
\date{}

\maketitle
\thispagestyle{empty}

\begin{abstract}
The $\Z_{2^s}$-additive codes are subgroups of $\Z^n_{2^s}$, and can be seen as a generalization of linear codes over $\Z_2$ and $\Z_4$. A
$\Z_{2^s}$-linear code is a binary code which is the Gray map image of a $\Z_{2^s}$-additive code. We consider $\Z_{2^s}$-additive simplex codes of type $\alpha$ and $\beta$, which are a generalization over $\Z_{2^s}$ of the binary simplex codes. These $\Z_{2^s}$-additive simplex codes are related to the $\Z_{2^s}$-additive Hadamard codes. In this paper, we use this relationship to establish the kernel of their binary images, under the Gray map, the $\Z_{2^s}$-linear simplex codes. Similar results can be obtained for the binary Gray map image of $\Z_{2^s}$-additive MacDonald codes.  
\end{abstract}

\section{Introduction}

Let $\Z_{2^s}$ be the ring of integers modulo $2^s$ with $s\geq1$. The set of
$n$-tuples or vectors over $\Z_{2^s}$ is denoted by $\Z_{2^s}^n$. A binary code of length $n$ is a nonempty subset of $\Z_2^n$,
and it is linear if it is a subspace of $\Z_{2}^n$. A nonempty
subset of $\Z_{2^s}^n$ is a $\Z_{2^s}$-additive code if it is a subgroup of $\Z_{2^s}^n$.
Note that, when $s=1$, a $\Z_{2^s}$-additive code is a binary linear code and, when $s=2$,
it is a quaternary linear code or a linear code over $\Z_4$.

Let $\cS_n$ be the symmetric group of permutations on the set $\{1,\dots,n\}$.
Two binary codes, $C_1$ and $C_2$, are said to be equivalent if there is a vector $\textbf{a}\in \Z_2^n$ and a
permutation of coordinates $\pi\in \cS_n$ such that $C_2=\{ \textbf{a}+\pi(\textbf{c}) : \textbf{c} \in C_1 \}$.
Two $\Z_{2^s}$-additive codes, $\C_1$ and $\C_2$, are said to be permutation equivalent if they differ
only by a permutation of coordinates, that is, if there is a permutation of coordinates $\pi\in\cS_n$
such that $\C_2=\{ \pi(\textbf{c}) : \textbf{c} \in \C_1 \}$.

The Hamming weight of $\textbf{u}\in\Z_{2}^n$, denoted by $\wt_H(\textbf{u})$, is
the number of nonzero coordinates of $\textbf{u}$. The Hamming distance of
$\textbf{u},\textbf{v}\in\Z_{2}^n$, denoted by $d_H(\textbf{u},\textbf{v})$, is the number of
coordinates in which they differ.  Note that $d_H(\textbf{u},\textbf{v})=\wt_H(\textbf{v}-\textbf{u})$. The minimum distance of a
binary code $C$ is $d(C)=\min \{ d_H(\textbf{u},\textbf{v}) : \textbf{u},\textbf{v} \in C, \textbf{u} \not = \textbf{v}  \}$. The Lee
weight of an element $i\in\Z_{2^s}$ is $\wt_L(i)=\min\lbrace i, 2^s-i\rbrace$ and the Lee weight of a
vector $\textbf{u}=(u_1,u_2,\dots,u_n)\in\Z_{2^s}^n$ is
$\wt_L(\textbf{u})=\sum_{j=1}^n \wt_L(u_j)\in\Z_{2^s}$. The Lee distance of two
vectors $\textbf{u},\textbf{v}\in\Z_{2^s}^n$ is $d_L(\textbf{u},\textbf{v})=\wt_L(\textbf{v}-\textbf{u})$. The
minimum distance of a $\Z_{2^s}$-additive code $\C$ is $d(\C)=\min \{
d_L(\textbf{u},\textbf{v}) : \textbf{u},\textbf{v}\in\C, \textbf{u} \not = \textbf{v}  \}$.

In \cite{Sole}, the Gray map  from $\Z_4$ to $\Z_2^2$ is defined as
$\phi(0)=(0,0)$, $\phi(1)=(0,1)$, $\phi(2)=(1,1)$ and $\phi(3)=(1,0)$. There exist
different generalizations of this Gray map, which go from $\Z_{2^s}$ to
$\Z_2^{2^{s-1}}$ \cite{Carlet,dougherty,Krotov:2007}. The
one given in \cite{Carlet} by Carlet is the map
$\phi:\Z_{2^s}\rightarrow\Z_2^{2^{s-1}}$ defined as follows:
\begin{equation}\label{genGraymap}
\phi(u)=(u_{s-1},\dots,u_{s-1})+(u_0,\dots,u_{s-2})Y,
\end{equation}
where $u\in\Z_{2^s}$, $[u_0,u_1, \ldots, u_{s-1}]_2$ is the binary expansion of $u$, that is $u=\sum_{i=0}^{s-1}2^{i}u_i$ ($u_i \in \lbrace0,1\rbrace$), and $Y$ is a matrix of size $(s-1)\times2^{s-1}$ which columns are the elements of $\Z_2^{s-1}$.
In \cite{Krotov:2007}, the generalized Gray maps are defined in terms of a Hadamard code.
 Note that the rows of $Y$ form a basis of a first order Reed-Muller code, which is a linear Hadamard code,
 after adding the all-one row. Therefore, the Carlet's Gray map $\phi$
 is a particular case of the Gray maps from \cite{Krotov:2007}, satisfying that $\sum \lambda_i \phi(2^i) =\phi(\sum \lambda_i 2^i)$.
 In this paper, we focus on this Gray map $\phi$, and we define $\Phi:\Z_{2^s}^ n\rightarrow\Z_2^{n2^{s-1}}$ as the component-wise Gray map $\phi$.

Let $\C$ be a $\Z_{2^s}$-additive code of length $n$. We say that its binary image
$C=\Phi(\C)$ is a $\Z_{2^s}$-linear code of length $2^{s-1}n$.
Since $\C$ is a subgroup of
$\Z_{2^s}^n$, it is isomorphic to an abelian structure
$\Z_{2^s}^{t_1}\times\Z_{2^{s-1}}^{t_2}\times
\dots\times\Z_4^{t_{s-1}}\times\Z_2^{t_s}$, and we say that $\C$, or equivalently
$C=\Phi(\C)$, is of type $(n;t_1,\dots,t_{s})$.
Note that $|\C|=2^{st_1}2^{(s-1)t_2}\cdots2^{t_s}$.
Unlike linear codes over finite fields,
linear codes over rings do not have a basis, but there
exists a generator matrix for these codes. If $\C$ is a
$\Z_{2^s}$-additive code of type $(n;t_1,\dots,t_s)$, then a generator matrix
of $\C$ with minimum number of rows has exactly $t_1+\cdots+t_s$ rows. A $2$-linear combination of the elements of $\mathcal{B}=\{\mathbf{b}_1,\dots,\mathbf{b}_r\}\subseteq \Z_{2^s}^n$ is $\sum_{i=1}^{r}\lambda_i\mathbf{b}_i$, for $\lambda_i\in\Z_2$. We say that $\mathcal{B}$ is a $2$-basis of $\C$ if the elements in $\mathcal{B}$ are 2-linearly independent and any $\mathbf{c}\in\C$ is a $2$-linear combination of the elements of  $\cal{B}$.

Two structural properties of binary codes are the rank and dimension of the kernel. The rank of a binary code $C$ is simply the
dimension of the linear span, $\langle C \rangle$,  of $C$.
The kernel of a binary code $C$ is defined as
$\mathrm{K}(C)=\{\textbf{x}\in \Z_2^n : \textbf{x}+C=C \}$ \cite{BGH83}. If the all-zero vector belongs to $C$,
then $\mathrm{K}(C)$ is a linear subcode of $C$.
Note also that if $C$ is linear, then $K(C)=C=\langle C \rangle$.
We denote the rank of a binary code $C$ as $\rank(C)$ and the dimension of the kernel as $\kernel(C)$.
These invariants can be used to distinguish between nonequivalent binary codes, since equivalent ones
have the same rank and dimension of the kernel.

A binary code of length $n$, $2n$ codewords and minimum distance $n/2$ is called a (binary) Hadamard code.
Hadamard codes can be constructed from Hadamard matrices \cite{Key,WMcwill}.
The $\Z_{2^s}$-additive codes that, under the Gray map $\Phi$, give a
Hadamard code are called $\Z_{2^s}$-additive Hadamard codes and the
corresponding binary images are called $\Z_{2^s}$-linear
Hadamard codes \cite{KernelZ2s,Krotov:2007}. The classification of $\Z_{2^s}$-linear
Hadamard codes have been studied by using their rank and dimension of the kernel \cite{KernelZ2s,Rank}, and also by establishing some equivalences among some of them \cite{EquivZ2s}.

Binary simplex codes are the dual of the well-known binary Hamming codes. They can also be seen as shortened binary linear Hadamard codes, that is, the codes that consist of the codewords of a Hadamard code having 0 in the first coordinate and deleting this coordinate.
As generalizations of the binary simplex codes, $\Z_{2^s}$-additive simplex codes of type $\alpha$ and $\beta$ were introduced and studied in  \cite{Gupta-thesis,Gupta-paper}. Specifically, for $s=2$, the linearity of the Gray map image of these codes was determined.
In this paper, in order to study the linearity and  structure of the Gray map image of $\Z_{2^s}$-additive simplex codes for $s>2$, we relate them with the $\Z_{2^s}$-additive Hadamard codes. We use this relationship to find the kernel and its dimension. In Section \ref{sec:construction}, we recall the construction of $\Z_{2^s}$-additive simplex codes of type $\alpha$ and $\beta$; and describe their relationship with the $\Z_{2^s}$-additive  Hadamard codes.
In Section \ref{sec:kernel}, we use this relation to obtain the kernel of both families of $\Z_{2^s}$-additive simplex codes.
Finally, in Section \ref{sec:conclusion}, we give some conclusions and further research on this topic.

\section{Construction of $\Z_{2^s}$-additive simplex codes}
\label{sec:construction}

In this section, we describe the construction of the $\Z_{2^s}$-additive simplex codes of type $\alpha$ and $\beta$
presented in \cite{Gupta-thesis,Gupta-paper}, and establish the relationship of these codes with the $\Z_{2^s}$-additive Hadamard codes constructed in \cite{KernelZ2s,Krotov:2007}.

Let $G_k^\alpha$ be a $k\times 2^{sk}$ matrix over $\Z_{2^s}$ consisting of all possible distinct columns.
This matrix can be constructed inductively from
$$
G_1^\alpha =\left(\begin{array}{ccccc}
0 & 1 & 2 & \cdots & 2^s-1
\end{array}\right)
$$
and
\begin{equation*}
G_k^\alpha=\left(\begin{array}{ccccc}
\zero & \one & \two & \cdots & \mathbf{2^s-1}\\
G^\alpha_{k-1} & G^\alpha_{k-1} & G^\alpha_{k-1} & \cdots &  G^\alpha_{k-1}\\
\end{array}\right)
\end{equation*}
for $k\geq 2$,  where $\mathbf{0}, \mathbf{1},\mathbf{2},\ldots, \mathbf{2^{s}-1}$ are the vectors having the elements $0, 1, 2, \ldots, 2^s-1$ from $\Z_{2^s}$  in all its coordinates, respectively. The code generated by $G^\alpha_k$, denoted by $S_k^{\alpha}$, is called $\Z_{2^s}$-additive simplex code of type $\alpha$ with $k$ generators.

The $\Z_{2^s}$-additive simplex codes of type $\beta$ can also be constructed inductively as follows.
Consider the matrices,
\begin{equation*}
G_2^\beta=\left(\begin{array}{ccccccccccc}
\one & 0 & 2 & \cdots & 2^s-2\\
0\,1\,2\,\dots\,2^s-1 & 1 & 1 & \cdots & 1
\end{array}\right)
\end{equation*}
and 
\begin{equation*}
G_k^\beta=\left(\begin{array}{ccccccccccc}
\one & \zero & \two & \cdots & \mathbf{2^s-2}\\
G^\alpha_{k-1} & G^\beta_{k-1} & G^\beta_{k-1} & \cdots & G^\beta_{k-1}
\end{array}\right)
\end{equation*}
for $k\geq 3$.
Then, the code generated by $G_k^\beta$ is denoted by $S_k^\beta$ and is called $\Z_{2^s}$-additive simplex code of type $\beta$ with $k$ generators.


The construction of both families of $\Z_{2^s}$-additive simplex codes, of type $\alpha$ and $\beta$, is related to the construction of  $\Z_{2^s}$-additive Hadamard codes
 \cite{KernelZ2s,Krotov:2007}.
 Now, we recall the recursive construction of these codes, given in \cite{KernelZ2s}.
Let $A^{1,0,\dots,0}=(1)$. Then, if
we have a matrix $A=A^{t_1,\dots,t_s}$, for any $i\in \{1,\ldots,s\}$, we may construct the matrix
\begin{equation}
A_i=
\left(\begin{array}{cccc}
0\cdot \mathbf{2^{i-1}}  & 1\cdot \mathbf{2^{i-1}} & \cdots & (2^{s-i+1}-1)\cdot \mathbf{2^{i-1}}  \\
A & A &\cdots & A \\
\end{array}\right).
\end{equation}
Now, $A^{t'_1,\ldots,t'_s}=A_i$, where $t'_j=t_j$ for $j\not=i$ and $t'_i=t_i+1$. Finally, the code  generated by $A^{t_1,t_2,\dots,t_s}$ and denoted by  $\mathcal{H}^{t_1,t_2,\dots,t_s}$ is the $\Z_{2^s}$-additive Hadamard code of type $(n;t_1,t_2,\dots,t_s)$, where $n$ is the length of the code. Along this paper, we consider that the matrices $A^{t_1,t_2,\ldots,t_s}$ are constructed recursively starting from $A^{1,0,\ldots,0}$ in the following way. First, we add $t_1-1$ rows of order $2^s$, up to obtain $A^{t_1,0,\ldots,0}$; then $t_2$ rows of order $2^{s-1}$ up to generate $A^{t_1,t_2,0,\ldots,0}$; and so on, until we add $t_s$ rows of order $2$ to achieve $A^{t_1,t_2,\ldots,t_s}$.


\begin{theorem}\label{Theorem:punctured}
Let $k\in\mathbb{N}$, $k\geq 1$. Then, we have that
$$ G^\alpha_k = \tilde{A}^{k+1,0,\dots,0}, $$
where $\tilde{A}^{k+1,0,\dots,0}$ is the matrix $A^{k+1,0,\dots,0}$ after removing the last row, that is, the all-one row.
\end{theorem}

\begin{proof}
Straightforward from definitions of $G^\alpha_k$ and $A^{k+1,0,\dots,0}$.
\end{proof}


\begin{theorem}\label{Theorem:beta}
Let $k\in\mathbb{N}$, $k\geq 2$. Then, we have that
\begin{equation} \label{eq:Gbeta}
G_k^\beta=\left(\begin{array}{ccccc}
\smash{\raisebox{-.5\normalbaselineskip}{$A_r^{k,0,\dots,0}$}} & \zero & \two & \cdots & \mathbf{2^s-2}\\
                  & G_{k-1}^\beta & G_{k-1}^\beta & \cdots & G_{k-1}^\beta
\end{array}\right),
\end{equation}
where $A_r^{k,0,\dots,0}$ is the matrix $A^{k,0,\dots,0}$ after moving the last row, that is, the all-one row, to the top of the matrix.
\end{theorem}

\begin{example}\label{Example:GenMat}
Let $s=2$. Then, we have that
\begin{equation*}
G_2^\alpha=\left(\begin{array}{cccc}
0000&1111&2222&3333\\
0123&0123&0123&0123\\
\end{array}\right)
\end{equation*}
and
\begin{equation*}
G_3^\beta=\left(\begin{array}{cccccc}
1111&1111&1111&1111&000000&222222\\
0000&1111&2222&3333&111102&111102\\
0123&0123&0123&0123&012311&012311\\
\end{array}\right)
\end{equation*}
are generator matrices for the $\Z_4$-additive simplex codes, $S_2^\alpha$ and $S_3^\beta$, respectively.
We also have that
\begin{equation*}
A^{3,0}=\left(\begin{array}{cccc}
0000&1111&2222&3333\\
0123&0123&0123&0123\\
1111&1111&1111&1111\\
\end{array}\right)
\end{equation*}
is a generator matrix for the $\Z_4$-additive Hadamard code  $\mathcal{H}^{3,0}$.
Note that $G_2^\alpha = \tilde{A}^{3,0}$ and
$$G_3^\beta=\left(\begin{array}{ccc}
\smash{\raisebox{-.5\normalbaselineskip}{$A_r^{3,0}$}} & \zero & \two \\
                  & G_2^\beta & G_2^\beta
\end{array}\right).$$
\end{example}

\section{Kernel of $\Z_{2^s}$-linear simplex codes}
\label{sec:kernel}

In this section, we determine the kernel, and its dimension, of the Gray map image of the $\Z_{2^s}$-additive simplex codes of both types, $\alpha$ and $\beta$.
In the proofs, we use the relationship between these codes and the $\Z_{2^s}$-additive Hadamard codes of type $(n;k,0,\ldots,0)$, described in Section \ref{sec:construction}. Moreover, we also use the known results on the kernel of the Gray map image of the $\Z_{2^s}$-additive Hadamard codes,
established in \cite{KernelZ2s}.


\begin{lemma} \label{coroHadamard} \cite{KernelZ2s}
Let $\mathcal{H}^{k,0,\dots,0}$ be a $\Z_{2^s}$-additive Hadamard code with $k\geq2$. Let $\mathcal{H}_b$ be the subcode of $\mathcal{H}^{k,0,\dots,0}$ which contains all the codewords of order at most two. Then,
$$K(\Phi(\mathcal{H}^{k,0,\dots,0}))=\left\langle\Phi(\mathcal{H}_b), \Phi(\sum_{i=0}^{s-2}\mathbf{2^i})\right\rangle$$
and $\kernel(\Phi(\mathcal{H}^{k,0,\dots,0}))=k+1$.
\end{lemma}

\begin{lemma} \label{addition} 
\cite{TapVeg:2003}
Let $u,v\in\Z_{2^s}$. Then, $\phi(u)+\phi(v)=\phi(u+v-2(u\odot v))$.
\end{lemma}

\begin{corollary}\label{coro:2s-1} \cite{KernelZ2s,TapVeg:2003}
Let $u\in\Z_{2^s}$. Then, $\phi(u)+\phi(2^{s-1})=\phi(u+2^{s-1})$.
\end{corollary}



\begin{lemma} \label{lemm:CondInKernel}
Let $\C$ be a $\Z_{2^s}$-additive code and ${\bf u}\in\C$. Then, $\Phi({\bf u})\in K(\Phi(\C))$ if and only if $2({\bf u}\odot {\bf v})\in\C$ for all ${\bf v}\in\C$.
\end{lemma}

\begin{proof}
We have that $\Phi({\bf u})+\Phi({\bf v})=\Phi({\bf u}+{\bf v}-2({\bf u}\odot {\bf v}))$  for any two vectors ${\bf u}, {\bf v}$ over $\Z_{2^s}$, by Lemma \ref{addition}. Therefore, $\Phi({\bf u})+\Phi({\bf v})\in \Phi(\C)$ if and only if ${\bf u}+{\bf v}-2({\bf u}\odot {\bf v}) \in \C$.
If ${\bf u}, {\bf v}\in\C$, since $\C$ is $\Z_{2^s}$-additive, then  ${\bf u}+{\bf v}-2({\bf u}\odot {\bf v}) \in \C$ if and only if $2({\bf u}\odot {\bf v})\in \C$.

Let ${\bf u}\in\C$. We have that $\Phi({\bf u})\in K(\Phi(\C))$ if and only if $\Phi({\bf u}) + \Phi({\bf v})\in\Phi(\C)$ for all ${\bf v}\in\C$; that is, if and only if $2({\bf u}\odot {\bf v})\in \C$.
\end{proof}

\begin{lemma} \label{lem:incode2}
Let $S_1^\alpha$ be the $\Z_{2^s}$-additive simplex code of type $\alpha$, which is generated by the matrix $G_1^\alpha =\left(\begin{array}{ccccc}
0 & 1 & 2 & \cdots & 2^s-1
\end{array}\right)$. Let ${\bf c}\in S_1^\alpha$ of order $2^s$. If $s>2$, then
$2({\bf c} \odot 2^i{\bf c}) \notin S_1^\alpha$ for all $i\in \{1,\ldots,s-2\}$.
\end{lemma}

\begin{proof} Since ${\bf c}\in S_1^\alpha$ is a codeword of order $2^s$, then ${\bf c}=\lambda(0,1,2,\cdots,2^s-1)$ for
an odd $\lambda\in\Z_{2^s}$. Therefore, the coordinates of ${\bf c}$ contain all the elements of $\Z_{2^s}$.
Let $j$ be the coordinate of ${\bf c}$ such that $c_j=1$, and $k$ the coordinate such that $c_k=2^{s-1}-1$.
Note that the $j$th coordinate of all nonzero codewords of $S_1^\alpha$
is always different to $0$. Let ${\bf w}=2({\bf c} \odot 2^i{\bf c})$.
Since $c_j=1$ and $i\geq 1$, the $j$th coordinate of ${\bf w}$ is $2(1 \odot 2^i)=0$.
Finally, we show that the $k$th coordinate of ${\bf w}$ is nonzero.
The binary expansion of $c_k=2^{s-1}-1$ is $[1,\ldots,1,0]_2$,
so the binary expansion of $2^ic_k$ is $[0,\ldots,0,1\ldots,1]_2$ having zeros
in the first $i$ positions. Since $i\leq s-2$, we have that $c_k \odot 2^ic_k\neq 0$. Therefore, ${\bf w}$
has a zero in the $j$th coordinate and it is not the all-zero codeword, so ${\bf w}\not \in S_1^\alpha$.
\end{proof}

\begin{proposition} \label{prop:kernelAlpha1}
Let $(S_1^\alpha)_b$ be the subcode of $S_1^\alpha$ which contains all its codewords of order at most two,
that is, $(S_1^\alpha)_b = \{ \zero, (0,2^{s-1},0,2^{s-1},\ldots,0 ,2^{s-1})\}$.
Then,
\begin{description}
\item{(i)} if $s=2$, $K(\Phi(S^\alpha_1))= \Phi(S_1^\alpha)$ and $\ker(\Phi(S^\alpha_1)) = 2$;
\item{(ii)} if $s>2$, $K(\Phi(S^\alpha_1))= \Phi((S_1^\alpha)_b)$ and $\ker(\Phi(S^\alpha_1)) = 1$.
\end{description}
\end{proposition}

\begin{proof}
For $s=2$, it is easy to check that the result is true.
Now, we assume that $s>2$. Clearly, $\Phi((S_1^\alpha)_b) \subseteq K(\Phi(S^\alpha_1))$ by Corollary \ref{coro:2s-1}. Let ${\bf c}\in S_1^\alpha\setminus (S_1^\alpha)_b$.
On the one hand, if ${\bf c}$ is of order $2^s$, then $2({\bf c} \odot 2{\bf c}) \notin S_1^\alpha$ by Lemma \ref{lem:incode2}.
On the other hand, if ${\bf c}$ is of order $2^{s-i}$, $i \in \{1,\ldots,s-2\}$, there is a codeword ${\bf c}' \in S_1^\alpha$ of order $2^s$ such that
$2^i {\bf c}'={\bf c}$. Again, by Lemma \ref{lem:incode2}, we have that $2({\bf c}'\odot {\bf c}) \not \in S_1^\alpha$.  In both cases, we have that $\Phi({\bf c}) \notin K(S_1^\alpha)$ by Lemma~\ref{lemm:CondInKernel}. Finally, since $(S_1^\alpha)_b = \{ \zero, (0,2^{s-1},0,2^{s-1},\ldots,0 ,2^{s-1})\}$, $\ker(\Phi(S^\alpha_1)) = 1$.
\end{proof}

\medskip
If $C$ is a code of length $n$ and $I\subseteq \{1,\ldots,n\}$, we denote $C_I=\{c_I : c \in C\}$, where $c_I$ denote the
restriction of the codeword $c$ to the coordinate positions in $I$.

\begin{theorem}\label{Theorem:KerAlpha}
Let $k\in\mathbb{N}$, $k\ge 1$. Let $(S_k^\alpha)_b$ be the subcode of $S_k^\alpha$ which contains all its codewords of order at most two.
Then, we have that
\begin{description}
	\item{(i)} if $k=1$ and $s=2$, then $K(\Phi(S^\alpha_1))= \Phi(S_1^\alpha)$ and $\ker(\Phi(S^\alpha_1)) = 2$;
	\item{(ii)} otherwise, $K(\Phi(S^\alpha_k))= \Phi((S_k^\alpha)_b)$ and $
	\ker(\Phi(S^\alpha_k))=k.$
\end{description}
\end{theorem}

\begin{proof}
If $k=1$, then the result follows from Proposition \ref{prop:kernelAlpha1}. Assume that $k>1$ and $s\geq 2$.
By construction, $(S^\alpha_k)_I = {\mathcal H}^{k,0,\ldots,0}$,
where $I=\{(2^s)^{k-1}+1,\ldots, 2\cdot (2^s)^{k-1}\}\subseteq  \{1,\ldots, 2^{sk}\}$. Therefore,  $K(\Phi(S^\alpha_k))_{\phi(I)}
= K(\Phi({\mathcal H}^{k,0,\ldots,0}))$, where $\phi(I)=\cup_{j \in I}\{2^{s-1}(j-1)+1, \dots,2^{s-1}j \}$, that is, $\phi(I)$ is the set of the corresponding coordinate positions of $I$ after applying the Gray map. By Lemma \ref{coroHadamard},
we have that $K(\Phi(\mathcal{H}^{k,0,\dots,0}))=\left\langle\Phi(\mathcal{H}_b), \Phi(\sum_{i=0}^{s-2}\mathbf{2^i})\right\rangle$,
where $\mathcal{H}_b$ contains the codewords of order at most two in $\mathcal{H}^{k,0,\dots,0}$.
We also have that $\Phi((S_k^\alpha)_b) \subseteq K(\Phi(S^\alpha_k))$ by Corollary \ref{coro:2s-1}. Moreover,
it is easy to see that $\Phi((S_k^\alpha)_b)_{\phi(I)} = \Phi(\mathcal{H}_b)$.

Assume that $K(\Phi(S^\alpha_k)) \not = \Phi((S_k^\alpha)_b)$.
Then, there exists ${\bf c} \in S^\alpha_k$
such that ${\bf c}_I=(\sum_{i=0}^{s-2}{\mathbf 2^i})$ and $\Phi({\bf c}) \in K(\Phi(S_k^\alpha))$. By the definition of $S_k^\alpha$,
${\bf c}= (\sum_{i=0}^{s-2} 2^i) ({\bf 0}, {\bf 1}, \ldots, {\bf 2^s-1})$.
Thus, if we prove that $\Phi({\bf c}) \notin K(\Phi(S_k^\alpha))$, then we obtain a contradiction and $K(\Phi(S^\alpha_k))= \Phi((S_k^\alpha)_b)$,
  so $\ker(\Phi(S^\alpha_k))=k$.

The codeword $\bf c$ induces a partition of all coordinate positions into $2^s$ blocks of $(2^s)^{k-1}$ consecutive coordinate positions,
such that all the coordinates of $\bf c$ in the same block are equal.
Since $\sum_{i=0}^{s-2}2^i$ is always an odd number, there are two blocks of coordinate positions, $I_1$ and $I_2$, such that ${\bf c}_{I_1}={\bf 1}$
and ${\bf c}_{I_2}={\bf 2^{s-1}-1}$. We consider the codeword $2{\bf c} \in S_k^\alpha$.
Note that, by construction, if a codeword of $S_k^\alpha$ had at least two blocks with all zeros,
then it would be the all-zero codeword.
Then, by the same argument as in the proof of Lemma \ref{lem:incode2}, we have that ${\bf w}=2({\bf c} \odot 2{\bf c}) \notin S_k^\alpha$.
Therefore, $\Phi({\bf c}) \notin K(\Phi(S_k^\alpha))$ by Lemma \ref{lemm:CondInKernel}.
\end{proof}

\medskip
Note that a direct consequence of Theorem \ref{Theorem:punctured} is that $K(\Phi(S^\alpha_k))\subset K(\Phi(\mathcal{H}^{k+1,0,\dots,0}))$.

\begin{corollary}
The binary $\Z_{2^s}$-linear simplex code $\Phi(S_k^\alpha)$ is nonlinear for all $s\geq 2$, $k\ge 1$; 
except for $s=2$ and $k=1$.
\end{corollary}

\begin{theorem}\label{Theorem:KerBeta}
Let $k\in\mathbb{N}$, $k\geq 2$. Let $(S_k^\beta)_b$ be the subcode of $S_k^\beta$ which contains all its codewords of order at most two.
Then, we have that
$$
K(\Phi(S^\beta_k))= \Phi((S_k^\beta)_b)
\quad \textrm{and} \quad
\ker(\Phi(S^\beta_k))=k.$$
\end{theorem}

\begin{proof}
By construction, $(S_k^\beta)_{I_1}={\mathcal H}^{k,0,\ldots,0}$, where $I_1=\{1,\ldots, (2^s)^{k-1}\}$.
Let ${\bf w}_1$ and ${\bf w}_2$ be the first and second row of the generator matrix $G_k^\beta$ given in (\ref{eq:Gbeta}), respectively.
By Lemma \ref{coroHadamard} and using the same arguments as in the proof of Theorem \ref{Theorem:KerAlpha},
we only need to prove that $\Phi({\bf c}) \notin K(\Phi(S_k^\beta))$,
where ${\bf c}= (\sum_{i=0}^{s-2} 2^i) {\bf w}_1$. By Lemma \ref{lemm:CondInKernel}, we just need to find a codeword ${\bf d} \in S_k^\beta$
such that $2({\bf c} \odot {\bf d}) \notin S_k^\beta$. In that case, we have that $K(\Phi(S^\beta_k))= \Phi((S_k^\beta)_b)$ and hence $\ker(\Phi(S^\beta_k))=k$.

We have that $S_k^\beta$ has length $n=2^{(s-1)(k-1)}(2^k-1)$ \cite{Gupta-paper}.
Thus, $\Phi(S_k^\beta)$ has length $2^{s-1}n$ and minimum Hamming distance $2^{s-2}n=2^{sk-k-1}(2^k-1)$.

If $k=2$, we consider the codeword ${\bf d}={\bf w}_2$. Since the binary expansion of $\sum_{i=0}^{s-2} 2^i$ is $[1,\ldots,1,0]_2$,
${\bf c}_{I_1} = (\sum_{i=0}^{s-2} 2^i) (1,1,\ldots, 1)$ and ${\bf d}_{I_1}=(0,1,\ldots,2^s-1)$, we have that $2({\bf c} \odot {\bf d})_{I_1}=2(0,1,\ldots, 2^{s-1}-1,0,1,\ldots,2^{s-1}-1)$. Let $I_2=\{1,\ldots, n\}\backslash I_1$. Since the coordinates of ${\bf c}_{I_2}$ are all even,
and the coordinates of ${\bf d}_{I_2}$ are all ones, we have that $2({\bf c} \odot {\bf d})_{I_2}=(0,\ldots,0)$.
Therefore, $\wt_H(\Phi(2({\bf c} \odot {\bf d})))=2^s\cdot 2^{s-2}=2^{2s-2}$. Finally, since all codewords of $\Phi(S_2^\beta)$
have Hamming weight $2^{2s-3}\cdot 3$, we obtain that $2({\bf c} \odot {\bf d})\notin S_k^\beta$.

If $k>2$, we consider the codeword ${\bf d}=2^{s-2}{\bf w}_2$. By construction, in the coordinates of $({\bf w}_2)_{I_1}$, each element of $\Z_{2^s}$ appears
the same number of times. If $[a_0,\ldots,a_{s-2},a_{s-1}]_2$ is the binary expansion of one of the coordinates, after multiplying by
$2^{s-2}$, it becomes $[0,\ldots,0,a_0,a_1]_2$. Again, since the binary expansion of $\sum_{i=0}^{s-2} 2^i$ is $[1,\ldots,1,0]_2$
and ${\bf c}_{I_1} = (\sum_{i=0}^{s-2} 2^i) (1,1,\ldots, 1)$,
we have that $2({\bf c} \odot {\bf d})_{I_1}$ contains half of the coordinates equal to $0$ and the other half equal to $2^{s-1}$.
Recall that $\wt_H(\phi(2^{s-1}))=2^{s-1}$.  Therefore, $\wt_H(\Phi(2({\bf c} \odot {\bf d})_{I_1}))=2^{s-1}\cdot (2^s)^{k-1}/2=2^{ks-2}$.

Let $I_2$ be the subset of $\{1,\ldots,n\}\backslash I_1$ such that $({\bf w}_2)_{I_2}={\bf 1}$,
and $I_3$ the subset such that $\{1,\ldots,n\}=I_1\cup I_2\cup I_3$. Since $\sum_{i=0}^{s-2} 2^i$ is always odd, by construction, in the coordinates of ${\bf c}_{I_2}$, each
element of the set $\{0,2,\ldots, 2^s-2\}$ appears the same number of times. Moreover, ${\bf c}_{I_2}$ induces a partition of all coordinate positions
of $I_2$ into $2^{s-1}$ blocks of $(2^s)^{k-2}$ consecutive coordinate positions, such that all the coordinates of ${\bf c}_{I_2}$ in the same
block are equal. For each $j\in \{0,1,\ldots,2^{s-1}-1\}$, there is a block of
coordinates $I_{2,j}$ such that ${\bf c}_{I_{2,j}}={\bf 2j}$. Let $[0,a_1,\ldots,a_{s-1}]_2$ be the binary expansion of $2j$.
The binary expansion of any coordinate in ${\bf d}_{I_{2}}={\bf 2^{s-2}}$ is $[0,\ldots,0,1,0]_2$.
On the one hand, if $s>2$, we have that
$2({\bf c} \odot {\bf d})_{I_2}$ contains in half of the blocks all coordinates equal to $0$, and the other half equal to $2^{s-1}$. That is, there are $2^{s-1}/{2}$ blocks, each one with $(2^s)^{k-2}$ coordinates equal to $2^{s-1}$.
Therefore, in this case, $\wt_H(\Phi(2({\bf c} \odot {\bf d})_{I_2}))=2^{s-1}\cdot 2^{s-1}/2 \cdot (2^s)^{k-2}=2^{ks-3}$.
On the other hand, if $s=2$, then $2({\bf c} \odot {\bf d})_{I_2}={\bf 0}$ and  $\wt_H(\Phi(2({\bf c} \odot {\bf d})_{I_2}))=0$.
For the coordinates in $I_3$, since all coordinates in ${\bf c}_{I_3}$ are even
and the ones in ${\bf d}_{I_3}$ are either $0$ or $2^{s-1}$, we have that $2({\bf c} \odot {\bf d})_{I_3}={\bf 0}$,
so $\wt_H(\Phi(2({\bf c} \odot {\bf d})_{I_3}))=0$.

Finally, if $s>2$, then $\wt_H(\Phi(2({\bf c} \odot {\bf d}))) = 2^{ks-2} + 2^{ks-3}=
2^{ks-3}\cdot 3$, which is not equal to the weight of all the codewords in $\Phi(S_k^\beta)$, so $2({\bf c} \odot {\bf d}) \notin S_k^\beta$.
If $s=2$, then $\wt_H(\Phi(2({\bf c} \odot {\bf d}))) = 2^{ks-2}$, and we also obtain that $2({\bf c} \odot {\bf d}) \notin S_k^\beta$.
\end{proof}

\begin{corollary}
	The binary  $\Z_{2^s}$-linear simplex code $\Phi(S_k^\beta)$ is nonlinear for all $s\geq 2$, $k\ge 2$.
\end{corollary}

\begin{example} Consider the generator matrices $G_2^\alpha$ and $G_3^\beta$ of the codes $S_2^\alpha$ and $S_3^\beta$, respectively, given in Example \ref{Example:GenMat}. The kernel of these codes  are
$$
K(\Phi(S_2^\alpha))=\begin{array}{c}
\langle\Phi(0000222200002222),\\
       \Phi(0202020202020202)\rangle\\
       \end{array}
$$
and
$$
K(\Phi(S_3^\beta))=\begin{array}{c}
\langle\Phi(2222222222222222000000000000),\\
\Phi(0000222200002222222200222200),\\
\Phi(0202020202020202020222020222)\rangle,\\
\end{array}
$$
and, it is clear that $\kernel(\Phi(S_2^\alpha))=2$ and $\kernel(\Phi(S_3^\beta))=3$.

Define $\C_2^\alpha$ and $\C_3^\beta$ the $\Z_4$-additive codes generated by $2G_2^\alpha$ and $2G_3^\beta$, respectively. By Corollary \ref{coro:2s-1}, we have that $\Phi(\C_2^\alpha)$ and $\Phi(\C_3^\beta)$ are linear. In fact, we have that $\C_2^\alpha=(S_2^\alpha)_b$ and $\C_3^\beta=(S_3^\beta)_b$ and hence $K(\Phi(S_2^\alpha))= \Phi(\C_2^\alpha)$ and $K(\Phi(S_3^\beta))=\Phi(\C_3^\beta)$.
\end{example}
%
%
%
%
%
%

Table \ref{table:RankAndKernel} shows the values of the invariants, rank and dimension of the kernel, for both kinds of
$\Z_{2^s}$-linear simplex codes and the related $\Z_{2^s}$-linear Hadamard codes. The rank has been computed by using \textsc{Magma} software \cite{Magma}. Some values in the table are missing because they are too hard to compute them or, in the case of $S_1^\beta$, because they do not exist. From this table, we can see that, as we show in Section \ref{sec:kernel}, the dimension of the kernel of both types of $\Z_{2^s}$-additive simplex codes are the same. Furthermore, we see that the rank is also the same, so we can conjecture that is like this for all $s,k\in\mathbb{N}$.

\begin{table}[h]
\caption{Rank and dimension of the kernel of $\Z_{2^s}$-linear Hadamard and simplex codes with $k$ generators for $1\leq k \leq 4$ and
$2\leq s\leq 4$.}
\label{table:RankAndKernel}
\begin{center}
\begin{tabular}{|c|c|c|c|c|c|}
\hline
                           &                         & $k=1$    & $k=2$    & $k=3$     & $k=4$    \\
\hline
\multirow{3}{*}{$\Z_4$}    & $\mathcal{H}^{k+1,0,0}$ & (3,8)  & (4,7)  & (5,11)  & (6,16) \\
\cline{2-6}
                           & $S_k^\alpha$                   & (2,2)  & (2,5)  & (3,9)   & (4,14) \\
\cline{2-6}
                           & $S_k^\beta$             & (-,-)  & (2,5)  & (3,9)   & (4,14) \\
\hline
\multirow{3}{*}{$\Z_8$}    & $\mathcal{H}^{k+1,0,0}$ & (3,8)  & (4,17) & (5,32)  & (6,56) \\
\cline{2-6}
                           & $S_k^\alpha$                   & (1,4)  & (2,12) & (3,26)  & (4,49) \\
\cline{2-6}
                           & $S_k^\beta$             & (-,-)  & (2,12) & (3,26)  & (4,49) \\
\hline
\multirow{3}{*}{$\Z_{16}$} & $\mathcal{H}^{k+1,0,0}$ & (3,14) & (4,44) & (5,121) & (6,-)  \\
\cline{2-6}
                           & $S_k^\alpha$                   & (1,7)  & (2,32) & (3,101) & (4,-)  \\
\cline{2-6}
                           & $S_k^\beta$             & (-,-)  & (2,32) & (3,101) & (4,-) \\
\hline
\end{tabular}
\end{center}
\end{table}

\section{Construction and kernel of $\Z_{2^s}$-linear MacDonald codes}

In this section, we determine the kernel, and its dimension,
of the Gray map image of the $\Z_{2^s}$-additive MacDonald codes of both types $\alpha$ and $\beta$
similarly as in Section \ref{sec:kernel} for $\Z_{2^s}$-additive simplex codes.
The $\Z_4$-additive MacDonald codes  were introduced in \cite{GuptaMacDonald}, and some properties were investigated in that paper.
These codes over $\Z_4$ give rise to some optimal two weight Hamming binary codes and codes meeting the Griesmer bound \cite{GuptaMacDonald}.

The $\Z_{2^s}$-additive MacDonald codes of type $\alpha$ and $\beta$ can be defined from the generator matrices
of the $\Z_{2^s}$-additive simplex codes of type $\alpha$ and $\beta$, respectively. In general, they can be defined in the same way as for
$s=2$ in \cite{GuptaMacDonald}.  For $1\leq u \leq k-1$, let
$G_{k,u}^\alpha$ be the matrix obtained from $G_k^\alpha$ by deleting columns corresponding to the columns of
$G_u^\alpha$, i.e.
\begin{equation} \label{mat:MDalpha}
G_{k,u}^\alpha=\left(\begin{array}{cc}
G_{k}^\alpha   & \backslash \quad \frac{{\mathbf 0}}{G_{u}^\alpha} \\
\end{array}\right),
\end{equation}
where $(A\backslash B)$ denotes the matrix obtained from the matrix $A$ by deleting the columns of the matrix $B$, and ${\mathbf 0}$ in (\ref{mat:MDalpha}) is a $(k-u) \times 2^{su}$ zero matrix.
Then, the $\Z_{2^s}$-additive MacDonald code $M_{k,u}^\alpha$ is the code generated by
$G_{k,u}^\alpha$.

\begin{proposition} \cite{GuptaMacDonald} Let $M_{k,u}^\alpha$ be the $\Z_{4}$-additive MacDonald code.
The Gray map image $\Phi(M_{k,u}^\alpha )$ is a nonlinear
$$(2^{sk+s-1}-2^{su+s-1}, 2^{sk}, 2^{sk+s-2}-2^{su+s-2})$$
binary two Hamming weight code with possible weights $2^{sk+s-2}-2^{su+s-2}$ and $2^{sk+s-2}$.
\end{proposition}

Similarly, for $1\leq u \leq k-1$, let
$G_{k,u}^\beta$ be the matrix obtained from $G_k^\beta$ by deleting columns corresponding to the columns of
$G_u^\beta$, i.e.
\begin{equation}\label{mat:MDbeta}
G_{k,u}^\beta=\left(\begin{array}{cc}
G_{k}^\beta   & \backslash \quad \frac{{\mathbf 0}}{G_{u}^\beta} \\
\end{array}\right),
\end{equation}
where ${\mathbf 0}$ in (\ref{mat:MDbeta}) is a zero matrix.
Then, the $\Z_{2^s}$-additive MacDonald code $M_{k,u}^\beta$ is the code generated by
$G_{k,u}^\beta$.

\begin{proposition} \cite{GuptaMacDonald} Let $M_{k,u}^\beta$ be the $\Z_{4}$-additive MacDonald code.
The Gray map image $\Phi(M_{k,u}^\beta )$ is a nonlinear
binary code.
\end{proposition}

\begin{theorem}\label{Theorem:KerMDAlpha}
Let $k\in\mathbb{N}$, $k\geq 2$. Let $(M_{k,u}^\alpha)_b$  ($(M_{k,u}^\beta)_b$) be the subcode of $M_{k,u}^\alpha$  ($M_{k,u}^\alpha$) which contains all its codewords of order at most two.
Then, we have that
$$
K(\Phi(M^\alpha_{k,u}))= \Phi((M_{k,u}^\alpha)_b)
\quad
\ker(\Phi(M^\alpha_{k,u}))=k$$
$$
K(\Phi(M^\beta_{k,u}))= \Phi((M_{k,u}^\beta)_b)
\quad \textrm{and} \quad
\ker(\Phi(M^\beta_{k,u}))=k.$$
\end{theorem}

\begin{corollary}
The binary  $\Z_{2^s}$-linear MacDonald code $\Phi(M_{k,u}^\alpha)$ ($\Phi(M_{k,u}^\beta)$) is nonlinear for all $s\geq 2$, $k\ge 2$, and $1\leq u \leq k-1$.
\end{corollary}

\section{Conclusions}
\label{sec:conclusion}

In this paper, we have established the kernel, and its dimension, of $\Z_{2^s}$-linear simplex codes for any $s\geq 2$. We have also seen that the $\Z_4$-linear code $\Phi(S_1^\alpha)$ is the only $\Z_{2^s}$-linear simplex codes for $s\geq 2$ that is linear.
Further research on this topic would be to determine the binary span of these codes, and its dimension, that is, the rank.
It would be also interesting to obtain similar results for the family of $\Z_{2^s}$-linear MacDonald codes, which are
also related to  $\Z_{2^s}$-linear Hadamard and simplex codes.

\end{document}